\documentclass[runningheads]{llncs}
\usepackage{graphicx}
\usepackage{hyperref}
\usepackage{booktabs}
\usepackage{sidecap}
\usepackage{multirow}
\usepackage{array}

\usepackage{amssymb}
\usepackage{pifont}
\newcommand{\cmark}{\ding{51}}%
\newcommand{\xmark}{\ding{55}}%

\usepackage{adjustbox} 
\newcolumntype{R}[2]{%
    >{\adjustbox{angle=#1,lap=\width-(#2)}\bgroup}%
    l%
    <{\egroup}%
}

\begin{document}
\title{Automated Extraction of Fine-Grained Standardized Product Information from Unstructured Multilingual Web Data}
\titlerunning{Extracting Product Information from Unstructured Web Data}
%

\author{
Alexander Flick\inst{1}\orcidID{0000-0001-5273-0679} \and
Sebastian Jäger\inst{1}\orcidID{0000-0001-9420-8571} \and 
Ivana Trajanovska\inst{1}\orcidID{0000-0003-0374-1210} \and \\
Felix Biessmann\inst{1,2}\orcidID{0000-0002-3422-1026}
}
%
\authorrunning{Flick et al.}
%
\institute{Berlin University of Applied Sciences and Technology \and
Einstein Center Digital Future, Berlin, Germany\\
\email{alexander.flick@bht-berlin.de}
}

\maketitle              
\begin{abstract}
Extracting structured information from unstructured data is one of the key challenges in modern information retrieval applications, including e-commerce.
%
%
Here, we demonstrate how recent advances in machine learning, combined with a recently published multilingual data set with standardized  fine-grained product category information, enable robust product attribute extraction in challenging transfer learning settings.
 Our models can reliably predict product attributes across online shops, languages, or both. Furthermore, we show that our models can be used to match product taxonomies between online retailers.

\end{abstract}
%
%

\section{Introduction}
Recent research achievements in the field of machine learning (ML) \cite{brinkmann_improving_2021,Peeters2020IntermediateTO} have the potential to improve automated information extraction in applications such as e-commerce. However, the translation of these ML innovations into real-world application scenarios is impeded by the lack of publicly available data sets.
Here we demonstrate that recent advances in ML can be translated into automated information extraction applications when leveraging carefully curated data. 
%
%
%
To better assess the contribution of this study, we first highlight some relevant data sets and methods that aim at the automated extraction of structured data in the field of e-commerce. 

\paragraph{Public E-commerce Data Sets}
\label{sec:data sets}
We summarize publicly e-commerce data sets used for the automated extraction of product information in Table~\ref{tab:ecommerce-datasets}. To leverage the potential of ML, large and diverse data sets that follow a fine-grained product taxonomy are favorable. A common and detailed taxonomy is the Global Product Classification (GPC) standard, which "classifies products by grouping them into categories based on their essential properties as well as their relationships to other products"\cite{GPC}. For example, multiple \emph{Brick}s (shirts and shorts) can belong to the same \emph{Family} (clothing) but are different \emph{Class}es (upper and lower body wear)\footnote{See the GPC Browser for more examples: \url{https://gpc-browser.gs1.org/}}.

%
\begin{table}[h]
    \caption{Comparison of e-commerce data sets used for product attribute extraction and classification.
    Column \emph{GPC} means whether or not the data set follows the GPC taxonomy.
    }
    \begin{tabular}{l|cccccl}
    \multirow{2}{*}{} & regular & \multicolumn{3}{c}{multi-}                                                           & \multirow{2}{*}{GPC} & \multirow{2}{*}{size} \\ \cline{3-5}
                      &     updated                             & lingual & shop & family &                                   &                       \\ \hline
        Farfetch product meta data \cite{Sigir22}        &\xmark &\xmark &\xmark &\xmark &\xmark & 400K \\
        Product details on Flipkart \cite{FlipkartProducts}                        &\xmark &\xmark &\xmark &\cmark &\xmark & 20K \\
        Amazon browse node classification \cite{AmazonMLChallenge}   &\xmark &\xmark &\xmark &\cmark &\xmark & 3M \\
        Amazon product-question answering \cite{rozen-etal-2021-answering}                &\xmark &\xmark &\xmark &\cmark &\xmark & 17.3GB \\
        Rakuten data challenge \cite{lin2018overview}                    &\xmark &\xmark &\xmark &\cmark &\xmark & 1M \\
        MAVE \cite{yang_mave_2022} &\xmark &\xmark &\xmark &\cmark &\xmark & 2.2M \\
        Innerwear from victoria's secret \& co \cite{Victoria_secret_Data} &\xmark &\xmark &\cmark &\xmark &\xmark & 600K  \\
        WDC-MWPD \cite{zhang2020mwpd2020} &\xmark &\xmark &\cmark &\xmark &\cmark & 16K \\
        WDC-25 gold standard \cite{WDC}  &\xmark &\xmark &\cmark &\cmark &\cmark & 24K \\
        GreenDB \cite{greenDB}                                            &\cmark &\cmark &\cmark &\cmark &\cmark & $>$576K 
    \end{tabular}
    \label{tab:ecommerce-datasets}
\end{table}

\paragraph{Multilingual Fine-Grained Product Classification}
\label{sec:related_work}
There are few recent studies investigating automated extraction of standardized product information in text corpora. Brinkmann et al. \cite{brinkmann_improving_2021} study how hierarchical product classification benefits from domain-specific language modeling. They report an improvement of 0.012 weighted F1 score by using schema.org product\footnote{Website: \url{https://schema.org/Product}} annotations for pre-training.
Peeters et al. \cite{peeters_cross-language_2022} study cross-language learning for entity matching and demonstrate that multilingual transformers outperform single-language models (German BERT) by 0.143 F1 when trained on a single language (German) and tested on multiple (German and English). Furthermore, using additional training data for the second language (English) improves the performance by another 0.038 weighted F1.

These studies highlight the potential of modern ML methods for automated product attribute extraction. In this work, we show that transfer learning helps to extract structured information (product category) from unstructured data (product name and description) and to find reliable taxonomy mappings.







\section{Experiments}
\label{sec:experiments}
We evaluate three transfer learning scenarios for product classification:

\begin{enumerate}
    \item \textbf{Language Transfer:} training on data of one language, test on other language data
    \item \textbf{Shop Transfer:} training on data of one shop, test on other shop data
    \item \textbf{Language and Shop Transfer:} training on data of one shop and one language, test on data of different shops and languages 
\end{enumerate}

Furthermore, we study whether ML methods can be used to find reliable taxonomy mappings. For this, we apply a model trained for a \emph{target taxonomy} to data that uses a \emph{source taxonomy}. For each source category, the majority of predicted target categories define the mapping from source to target taxonomy.


\paragraph{Data Sets} In our experiments, we use two data sets, the GreenDB \cite{jagerGreenDBProductbyProductSustainability2022} and the Farfetch data set \cite{Sigir22}. The GreenDB\footnote{We use GreenDB version 0.2.2 available at \url{https://zenodo.org/record/7225336}} is a multilingual data set covering 5 European shops with about 576k unique products of the 37 most important product categories following the GPC taxonomy. It covers categories from the GPC segments Clothing, Footwear, Personal Accessories, Home Appliances, Audio Visual/Photography, and Computing. A recent publication \cite{jagerGreenDBDatasetBenchmark2022} presents the GreenDB's high quality and usefulness for information extraction tasks. The Farfetch data set has about 400k unique products from a single shop. It does not follow a public taxonomy and covers only fashion products.

%

\paragraph{ML Model} The experiment implementation is based on autogluon's \cite{agmultimodaltext} TextPredictor and uses \emph{mDeBERTav3} \cite{DeBERTaV3} as the backbone model. For training, we use the GreenDB and apply Cleanlab \cite{northcutt2021confidentlearning} to find and remove miss-classified products (211 were found). Our models use the product's name and description to predict their product category. $model_{baseline}$ is trained on the entire GreenDB (all shops), $model_{ZaDE}$ on the German, $model_{ZaFR}$ on the French, and $model_{ZaALL}$ on the German, French, and English Zalando products contained in the GreenDB.

\begin{figure}    
    \centering
    \includegraphics[scale=0.12]{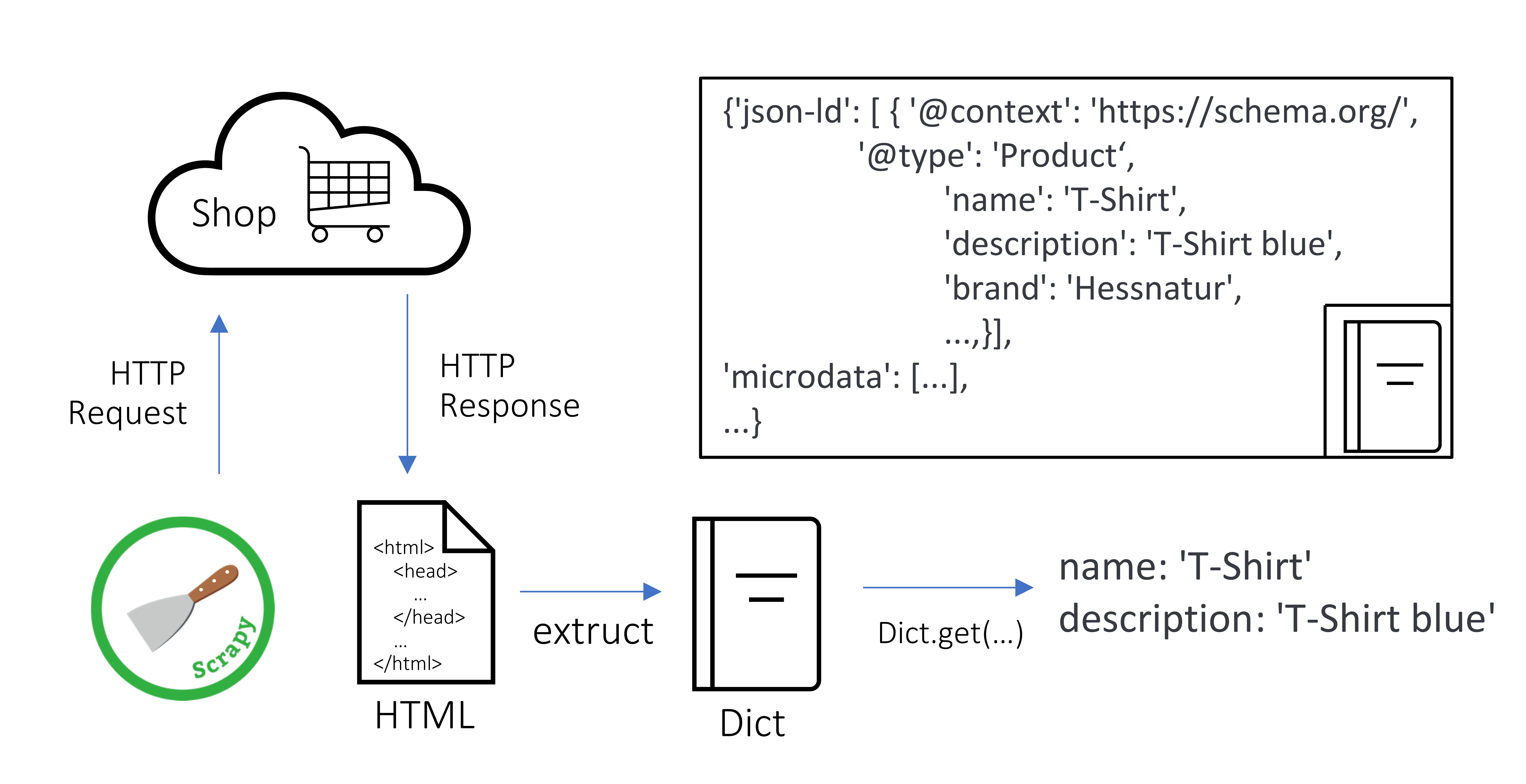}
    \caption{Online demo overview. Automated extraction of schema.org information (product name and description) from HTML, used for product classification.
    \label{fig:greendb-flowchart}
    }    
\end{figure}

\paragraph{Online Demo} To demonstrate the transfer capabilities, we published an online demo available: \url{https://product-classification.demo.calgo-lab.de}. As shown in Figure~\ref{fig:greendb-flowchart}, it automatically downloads the HTML of a given URL, extracts the products' name and description, and uses $model_{baseline}$ to predict its GPC category.

\section{Results}
The baseline performance ($model_{baseline}$) shows a strong 0.99 weighted F1 score on a GreenDB test set.

\paragraph{Transfer Tasks} $model_{ZaDE}$ demonstrates language transfer when it is applied to other languages of the same shop. It achieves weighted F1 scores of 0.898 for English and 0.873 for French.
Applying $model_{ZaFR}$ and $model_{ZaDE}$ on other shops demonstrates shop transfer with weighted F1 scores from 0.648 to 0.836. If the model is fine-tuned on multi-lingual data ($model_{ZaALL}$), almost all shops benefit, see Table~\ref{tab:shop_language_transfer} for details. The language and shop transfer is even more challenging and performs worse for all shops. Transferring across data sets, i.e., applying $model_{baseline}$ to Farfetch data, achieves a 0.924 weighted F1 score.




\begin{table}[h]
    \caption{Weighted F1 scores for shop transfer experiments. Scores from 0.648 to 0.836 demonstrate robust shop transfer. Shop transfer profits from additional data in other languages. }
    \centering
    \begin{tabular}{ll|cc|cc}
    \multirow{2}{*}{}                          & \multicolumn{1}{c|}{\multirow{2}{*}{Model}} & \multicolumn{2}{c|}{FR}                                & \multicolumn{2}{c}{DE}                                \\
                                               & \multicolumn{1}{c|}{}                       & Asos                      & H\&M                       & Otto                      & Amazon                      \\ \hline
    \multirow{3}{*}{Shop Transfer}             & $model_{ZaFR}$                                          & \multicolumn{1}{r}{0.836} & \multicolumn{1}{r|}{0.678} & -                         & -                         \\
                                               & $model_{ZaDE}$                                          & -                         & -                          & \multicolumn{1}{r}{0.777} & \multicolumn{1}{r}{0.648} \\
                                               & $model_{ZaALL}$                                         & \multicolumn{1}{r}{0.842} & \multicolumn{1}{r|}{0.717} & \multicolumn{1}{r}{0.762} & \multicolumn{1}{r}{0.739} \\ \hline
    \multirow{2}{*}{Shop \& Language Transfer} & $model_{ZaFR}$                                          & -                         & -                          & \multicolumn{1}{r}{0.614} & \multicolumn{1}{r}{0.449} \\
                                               & $model_{ZaDE}$                                          & \multicolumn{1}{r}{0.795} & \multicolumn{1}{r|}{0.666} & -                         & -
    \end{tabular}
    \label{tab:shop_language_transfer}
\end{table}

\paragraph{Taxonomy Matching} Using $model_{baseline}$ to map products' categories from Farfetch to GreenDB (GPC taxonomy) results in 41 out of 46 ($>$89\%) correctly mapped categories.

\section{Conclusion}
We demonstrate that combining rich multilingual data sets and modern ML methods enables fine-grained standardized product information extraction from unstructured data. We investigate several transfer learning settings when training and testing on data from different shops and languages, even in zero-shot scenarios when no data from another shop and language was available in the training data.
%



\clearpage
\noindent \textbf{Acknowledgements} This research was supported by the Federal Ministry for the Environment, Nature Conservation and Nuclear Safety based on a decision of the German Bundestag.

\scriptsize
\bibliographystyle{splncs04}
\bibliography{ECIR-2023.bib}
%
%
%
%

\end{document}